# Macroscopic Thermodynamics of Equilibrium Characterized by Power-Law Canonical Distributions

Sumiyoshi Abe[1] and A. K. Rajagopal[2]

[1]*College of Science and Technology, Nihon University,*

*Funabashi, Chiba 274-8501, Japan*

[2]*Naval Research Laboratory, Washington D. C. 20375-5320*

Macroscopic thermodynamics of equilibrium is constructed for systems obeying power-law canonical distributions. With this, the connection between macroscopic thermodynamics and microscopic statistical thermodynamics is generalized. This is complementary to the Gibbs theorem for the celebrated exponential canonical distributions of systems in contact with a heat bath. Thereby, a thermodynamic basis is provided for power-law phenonema ubiquitous in nature.

05.70.-a Thermodynamics

Statistical mechanics builds an essential bridge between the laws of nature governing microscopic dynamics of constituents of matter and its macroscopic behavior. One and only example of such a description known to date is the theory of Boltzmann and Gibbs characterized by the exponential distributions. Since many systems fall under the sway of this theory, there has been no attempt to find alternate possibilities. However, nowadays it is widely recognized that many phenomena in nature obey different kinds of distributions, i.e., the power-law distributions. We may cite here a few such examples: anomalous diffusion [1], vibrating powders [2], fully developed turbulence [3], and line shape cumulants in glasses [4]. We point out that even though these are typically nonequilibrium phenonema the structures of their distributions persist for inordinately long times so that it is natural to consider them as in quasi thermal equilibrium. In traditional Boltzmann-Gibbs theory, the canonical distribution function is derived from microcanonical ensemble with the principle of equal *a priori* probability (microscopic statistical concept) and is connected with macroscopic thermodynamic description by considering the system in contact with a macroscopic heat bath. That it is of crucial importance to establish the interrelationship between the two procedures was stressed by Tisza and Quay [5] and also by Kubo *et al.* [6].

Recently, it has been shown [7,8] that the power-law distribution can also be deduced based on microcanonical ensemble theory, thus providing a foundation to microscopic statistical thermodynamics of equilibrium in such a context. It has also been shown from the perspective of the (Lévy-Gnedenko) generalized central limit theorem [9] that this result is a natural extention of corresponding Boltzmann-Gibbs theory obtained more than half a century ago by Khinchin [10] using the ordinary central limit theorem. However, understanding of the power-law systems in a thermal equilibrium language is not complete until a way of realizing their macroscopic thermodynamic description is found.

The purpose of this paper is to present a macroscopic description of systems obeying the power-law distributions in equilibrium by reconsideration of contact with a heat bath. We establish the interrelationship between microscopic statistical and macroscopic thermal equilibria in parallel with the discussion of Boltzmann-Gibbs theory made in Ref. [5,6]. Thus, the Gibbsian measure is shown to be appropriately modified here. In this manner, a link between the laws on a microscopic scale and the macroscopic behavior of matter is extended to a much wider class of systems in nature.

We begin our discussion by considering a partition of energy between two systems, I and II, in thermal contact. These systems have energies $E_I$ and $E_{II}$, and the total energy is fixed as $E = E_I + E_{II}$. If the state densities of I and II are denoted by $\Omega_I$ and $\Omega_{II}$, respectively and that of the total system I+II by $\Omega$, then we have

$$\Omega(E)\Delta E = \Delta E \int_0^E \Omega_I(E_I)\, \Omega_{II}(E - E_I)\, dE_I \; , \qquad (1)$$

where $\Delta E$ is the width of the shell of constant energy in phase space of the total system. The probability of finding the system I in the range of the energy $(E_I, E_I + dE_I)$ is given by [6]

$$p(E_{\mathrm{I}})dE_{\mathrm{I}} = \frac{\Omega_{\mathrm{I}}(E_{\mathrm{I}})\,\Omega_{\mathrm{II}}(E - E_{\mathrm{I}})\Delta E}{\Omega(E)\Delta E}\,dE_{\mathrm{I}}. \tag{2}$$

From eq. (1), the normalization of this probability follows obviously. We are interested in the most probable partition of energy in conformity with the principle of equal *a priori* probability (microcanonical measure). This is determined by maximizing the quantity

$$\Omega_{\mathrm{I}}(E_{\mathrm{I}})\Omega_{\mathrm{II}}(E - E_{\mathrm{I}})\Delta E\,dE_{\mathrm{I}}. \tag{3}$$

In the traditional discussions of identifying the temperature of the system I and deriving its probability distribution, one analyzes the maximization procedure by the corresponding condition on the logarithm of the quantity in eq. (3). The underlying assumption hidden behind this treatment is extensivity (additivity) of entropy. To accomodate the power-law distributions of interest in the present work, we relax this assumption. From among many possible choices of accomplishing this, here we adopt, instead of the ordinary logarithmic operation mentioned above, the "*q*-logarithmic function" defined by [11]

$$\ln_q(x) = \frac{x^{1-q} - 1}{1 - q}, \tag{4}$$

where $q$ is a positive constant. This function has the following nonadditive property:

$$\ln_q(xy) = \ln_q(x) + \ln_q(y) + (1-q)\ln_q(x)\ln_q(y). \tag{5}$$

The inverse function associated with eq. (4) is the "$q$-exponential function"

$$e_q(x) = \begin{cases} [1+(1-q)x]^{1/(1-q)} & (1+(1-q)x > 0) \\ 0 & (1+(1-q)x \leq 0) \end{cases}, \tag{6}$$

that is, $e_q(\ln_q(x)) = x = \ln_q(e_q(x))$ for positive $x$. In the limit $q \to 1$, $\ln_q(x)$ and $e_q(x)$ converge to the ordinary $\ln(x)$ and $e^x$, respectively. An important point is that, just like $\ln(x)$, $\ln_q(x)$ is monotonically increasing concave function of $x$ for all positive values of $q$. (It is noted that this particular choice of the pair, $\ln_q(x)$ and $e_q(x)$, is one of many other possibilities.) Using the $q$-logarithmic operation, the maximization condition is now written as follows:

$$\ln_q\big(\Omega_{\mathrm{I}}(E_{\mathrm{I}})\,\Omega_{\mathrm{II}}(E-E_{\mathrm{I}})\big) = \max. \tag{7}$$

Vanishing of the first derivative with respect to $E_{\mathrm{I}}$ leads to

$$\frac{1}{1+(1-q)\ln_q(\Omega_{\mathrm{I}}(E_{\mathrm{I}}))}\frac{\partial \ln_q(\Omega_{\mathrm{I}}(E_{\mathrm{I}}))}{\partial E_{\mathrm{I}}}$$

$$= \frac{1}{1+(1-q)\ln_q(\Omega_{II}(E_{II}))} \frac{\partial \ln_q(\Omega_{II}(E_{II}))}{\partial E_{II}}, \tag{8}$$

where equation (5) has been used. If we identify

$$S_{i\,q} = S_q(\Omega_i) \equiv \ln_q(\Omega_i) \qquad (i = \text{I, II}) \tag{9}$$

as a generalized entropy and define a parameter $\beta_i$ by

$$\beta_i = \frac{\partial S_q(\Omega_i)}{\partial E_i}, \tag{10}$$

then we have

$$\frac{\beta_I}{c_{I\,q}} = \frac{\beta_{II}}{c_{II\,q}} \equiv \beta^*, \tag{11}$$

where we have introduced the notation

$$c_{i\,q} \equiv 1+(1-q)S_{i\,q} = \Omega_i^{1-q} \tag{12}$$

By analogy with the ordinary case $(q \to 1)$, the separation constant $\beta^*$ may be interpreted as the temperature of the total system. This can be thought of as the $q$-generalization of the zeroth law of thermodynamics. Similarly, the equilibrium conditions associated with the chemical potentials and the pressures can be obtained if the particle numbers and system volumes are respectively taken into account in the same manner as above.

If the two systems have different values of $q$, say $q_{\text{I}}$ and $q_{\text{II}}$, then the equilibrium condition in eq. (11) turns out to be modified to the following form:

$$\frac{\beta_{\text{I}}}{c_{\text{I}\,q_{\text{I}}}} = \frac{\beta_{\text{II}}}{c_{\text{II}\,q_{\text{II}}}} = \beta^*. \qquad (13)$$

Taking the limit $q_{\text{I}} \to 1$ in this equation, the system I is then regarded as the ordinary thermometer measuring the temperature of the system II. Thus, we identify $\beta^*$ with the ordinary inverse temperature $\beta$, which equates $\beta_{\text{II}}$ to $c_{\text{II}\,q_{\text{II}}}\beta$.

To ascertain the above state to be indeed the maximum we are seeking, we need to verify that the second derivative is negative. Carrying out such a calculation, we arrive at the following expression:

$$\frac{d^2}{dE_{\text{I}}^2}\ln_q\!\left(\Omega_{\text{I}}(E_{\text{I}})\Omega_{\text{II}}(E-E_{\text{I}})\right)$$

$$= \frac{c_{\text{II}\,q}}{c_{\text{I}\,q}}(1-q)\frac{d^2}{dE_{\text{I}}^2}\ln\!\left[1+(1-q)S_{\text{I}\,q}\right] + \frac{c_{\text{I}\,q}}{c_{\text{II}\,q}}(1-q)\frac{d^2}{dE_{\text{I}}^2}\ln\!\left[1+(1-q)S_{\text{II}\,q}\right], \qquad (14)$$

where it is noted that the ordinary logarithm appears on the right-hand side of this equation. As in the case of the discussion of ordinary Boltzmann-Gibbs theory, we also assume that the second-order derivative of the generalized entropy in eq. (9) is negative. Then, we see that each term on the right-hand side of eq. (14) is negative for all positive values of $q$, due to the monotonicity of the logarithmic function.

Now, let us study the necessary modification of the Gibbs theorem (for the canonical measure). For this purpose, we consider the system II to be a heat bath, and therefore

$$E_\mathrm{I} \ll E_\mathrm{II}. \tag{15}$$

In addition, we take the system I to be in its $k$th state of energy $E_\mathrm{I} = \varepsilon_k$. The probability of finding the system I in such a state is proportional to the number of microscopic states of the system II as follows [6]:

$$f(\varepsilon_k) \propto \frac{\Omega_\mathrm{II}(E - \varepsilon_k)}{\Omega_\mathrm{II}(E)}. \tag{16}$$

We rewrite this equation in the form

$$f(\varepsilon_k) \propto e_q\left(\ln_q \frac{\Omega_\mathrm{II}(E - \varepsilon_k)}{\Omega_\mathrm{II}(E)}\right). \tag{17}$$

Using the identity

$$\ln_q\left(\frac{x}{y}\right) = \frac{1}{y^{1-q}}\left[\ln_q(x) - \ln_q(y)\right], \tag{18}$$

we obtain

$$f(\varepsilon_k) \propto e_q\left(\frac{1}{\Omega_{\text{II}}^{1-q}(E)}\left[\ln_q(\Omega_{\text{II}}(E-\varepsilon_k)) - \ln_q(\Omega_{\text{II}}(E))\right]\right). \tag{19}$$

From eq. (15) with $E_{\text{I}} = \varepsilon_k$, we may perform the following expansion to the leading order of $\varepsilon_k$:

$$f(\varepsilon_k) \propto e_q\left(\frac{1}{\Omega_{\text{II}}^{1-q}(E)}\left[\ln_q(\Omega_{\text{II}}(E)) - \varepsilon_k \frac{\partial \ln_q(\Omega_{\text{II}}(E))}{\partial E} + \text{L} - \ln_q(\Omega_{\text{II}}(E))\right]\right)$$

$$\cong e_q(-\beta^* \varepsilon_k), \tag{20}$$

provided that equations (10) and (11) have been used. In the above, the second-order term has been neglected because of the assumed large size of the heat bath. Therefore, we have

$$\Omega_{\text{II}}(E - \varepsilon_k) \propto \Omega_{\text{II}}(E)\, e_q(-\beta^* \varepsilon_k). \tag{21}$$

This is the modification of the Gibbs measure promised earlier. For $q > 1$, $f(\varepsilon_k)$ in eq. (20) for large values of $\varepsilon_k$ obeys the power law

$$f(\varepsilon_k) \sim \varepsilon_k^{-1/(q-1)}, \tag{22}$$

as desired. On the other hand, for $0 < q < 1$, $f(\varepsilon_k)$ has a cut-off at $\varepsilon_{k\,\max} = \left[(1-q)\beta^*\right]^{-1}$ in view of eq. (6).

Since the number of degrees of freedom of the heat bath is assumed to be very large, it is appropriate to consider the relative probability of finding the system I as the ratio of the probability in the state with energy $\varepsilon_k$ relative to the fixed value of energy $\varepsilon_l$:

$$\pi(\varepsilon_k ; \varepsilon_l) = \frac{f(\varepsilon_k)}{f(\varepsilon_l)} = \frac{\Omega_{\mathrm{II}}(E-\varepsilon_k)}{\Omega_{\mathrm{II}}(E-\varepsilon_l)}, \tag{23}$$

which is rewritten as

$$\pi(\varepsilon_k ; \varepsilon_l) = e_q\!\left(\ln_q\!\left(\frac{\Omega_{\mathrm{II}}(E-\varepsilon_k)}{\Omega_{\mathrm{II}}(E-\varepsilon_l)}\right)\right)$$

$$= e_q\!\left(\frac{1}{\Omega_{\mathrm{II}}^{1-q}(E-\varepsilon_l)}\left[\ln_q\!\left(\Omega_{\mathrm{II}}(E-\varepsilon_k)\right) - \ln_q\!\left(\Omega_{\mathrm{II}}(E-\varepsilon_l)\right)\right]\right). \tag{24}$$

Performing the expansion with respect to $\varepsilon_k$ and $\varepsilon_l$ and keeping the leading order terms, we obtain

$$\pi(\varepsilon_k ; \varepsilon_l) = e_q\!\left(-\beta^*(\varepsilon_k - \varepsilon_l)\right). \tag{25}$$

Here, it should be noted that, to derive the relative probability of this form, we have to resort to the first principle calculation shown in eq. (24) and not to simply substitute the expression given in eq. (20) into the first part in eq. (23). This is a manifestation of the nonextensive (nonadditive) structure of the measure used here.

In conclusion, we have shown that macroscopic thermodynamics of equilibrium concerning two systems in contact can be realized in ways other than the well-known Gibbsian. This has been achieved by relaxing the assumption of extensivity of the measure in phase space in a completely consistent way. Thus, the connection between macroscopic thermodynamics and microscopic statistical thermodynamics is established for systems obeying power-law canonical distributions. It should also be noted that this demonstration is in conformity with Tsallis nonextensive statistical mechanics [11]. We stress that the particular choice of the pair of functions, $\ln_q(x)$ and $e_q(x)$, in the present work is just one among many other possibilities. Thus, nonuniqueness of canonical ensemble theory for macroscopic thermodynamics of equilibrium is revealed.

One of us (S.A.) was supported in part by the Grant-in-Aid for Scientific Research of Japan Society for the Promotion of Science and by the GAKUJUTSU-SHO Program of College of Science and Technology, Nihon University. The other (A. K. R.) acknowledges the partial support from the US Office of Naval Research. He also acknowledges the support from the International Academic Exchange Program of College of Science and Technology, Nihon University, which has enabled this collaboration. We would like to thank the referee for perceptive comments.**References**

[1] J. P. Bouchaud and A. Georges, Phys. Rep. **195**, 127 (1990).

[2] Y-H. Taguchi and H. Takayasu, Europhys. Lett. **30**, 499 (1995).

[3] C. Beck, Physica A **277**, 115 (2000).

[4] E. Barkai, R. Silbey, and G. Zumofen, Phys. Rev. Lett. **84**, 5339 (2000).

[5] L. Tisza and P. M. Quay, Ann. Phys. (NY) **25**, 48 (1963).

[6] R. Kubo, H. Ichimura, T. Usui, and N. Hashitsume, *Statistical Mechanics* (North Holland, Amsterdam, 1988).

[7] S. Abe and A. K. Rajagopal, Phys. Lett. A **272**, 341 (2000).

[8] S. Abe and A. K. Rajagopal, "Microcanonical " preprint (2000).

[9] S. Abe and A. K. Rajagopal, "Justification of power-law canonical distributions based on generalized central limit theorem " preprint (2000).

[10] A. I. Khinchin, *Mathematical Foundations of Statistical Mechanics* (Dover, New York, 1949)

[11] C. Tsallis, in *Nonextensive Statistical Mechanics and Its Applications*, eds. S. Abe and Y. Okamoto (Springer-Verlag, Heidelberg), to be published.12